\documentclass[11pt,prd,onecolumn,showpacs,amsmath,amssymb,aps,floats,floatfix,nofootinbib]{revtex4-1}
\usepackage[colorlinks=true,urlcolor=rossos,anchorcolor=black,citecolor=mygreen,filecolor=black,linkcolor=blue,menucolor=blue,linktocpage=true]{hyperref} % should be commented out if the tex file will be compiled with latex in arXiv!!! (pdflatex is fine)
\usepackage[multidot]{grffile}  % allow the name of figures to include dots
\usepackage{dcolumn}
\usepackage{bm}
\usepackage{amsmath}
\usepackage{amsfonts}
\usepackage{amssymb}
\usepackage{color}
\usepackage{footmisc}
\usepackage{latexsym}
\usepackage{slashed} % slash mark
\usepackage{pstricks}
\usepackage{indentfirst}
\usepackage{mathrsfs}
\usepackage{multirow}
\usepackage{epsfig,psfrag}
\usepackage{subfigure}
\usepackage{mathtools}
\usepackage{enumitem}
\usepackage{setspace} % spacing
\usepackage[utf8]{inputenc} % accept utf-8 input encoding
\usepackage[scientific-notation=true]{siunitx} % comprehensive units
\graphicspath{{figs/}}

\newcommand{\SUtwoL}{\mathrm{SU}(2)_\mathrm{L}}
\newcommand{\UoneY}{\mathrm{U}(1)_\mathrm{Y}}

\newcommand{\sw}{s_\mathrm{W}^2}
\newcommand{\cw}{c_\mathrm{W}^2}

\definecolor{rossos}{cmyk}{0,1,1,0.55}
\definecolor{mygreen}{rgb}{0.27, 0.64, 0.48}

\makeatother

\allowdisplaybreaks % allow eqnarray breaks
\begin{document}

\title{Electroweak dark matter model accounting for the CDF $W$-mass anomaly}

\author{Jin-Wei Wang$^{1,2,3}$}
\email{jinwei.wang@sissa.it}
\author{Xiao-Jun Bi$^{4,5}$}
\email{bixj@ihep.ac.cn}
\author{Peng-Fei Yin$^4$}
\email{yinpf@ihep.ac.cn}
\author{Zhao-Huan Yu$^6$}
\email{yuzhaoh5@mail.sysu.edu.cn}
 	
\affiliation{\vspace{0.4cm}$^1$Scuola Internazionale Superiore di Studi Avanzati (SISSA), via Bonomea 265, 34136 Trieste, Italy}
\affiliation{$^2$INFN, Sezione di Trieste, via Valerio 2, 34127 Trieste, Italy}
\affiliation{$^3$ Institute for Fundamental Physics of the Universe (IFPU), via Beirut 2, 34151 Trieste, Italy}
\affiliation{$^4$Key Laboratory of Particle Astrophysics, Institute of High Energy Physics, Chinese Academy of Sciences, Beijing 100049, China}
\affiliation{$^5$School of Physical Sciences, University of Chinese Academy of Sciences, Beijing 100049, China}
\affiliation{$^6$School of Physics, Sun Yat-Sen University, Guangzhou 510275, China}

\begin{abstract}
\large
\vspace{0.4cm}
Recently, the CDF collaboration reported a new measurement of the $W$ boson mass $M_W = 80.4335 \pm 0.0094$~GeV, which shows a $\sim 7\sigma$ deviation from the standard model prediction $80.3545 \pm 0.0057$~GeV obtained by the electroweak (EW) global fit. This deviation can be explained by new physics generating moderate EW oblique parameters $S$, $T$, and $U$. In this work, we use the loop corrections induced by some extra EW multiplets to explain the CDF $M_W$ anomaly. The lightest neutral particle in the multiplets can also serve as a candidate of cold dark matter (DM). We consider
two such models, namely singlet-triplet scalar DM and singlet-doublet fermionic DM models, and perform numerical scans to find the parameter points accounting for the $M_W$ anomaly. The constraints from the correct DM thermal relic density and direct detection are also taken into account. We find the parameter points simultaneously interpreting the $M_W$ anomaly and satisfying the DM requirements in the former model, but do not find such parameter points in the latter model.
\end{abstract}

%\pacs{13.66.Jn, 95.35.+d}
\maketitle
%\tableofcontents

\section{Introduction}
\label{sec:intro}

Electroweak (EW) precision observables are important and powerful probes for exploring the new physics (NP) beyond the standard model (SM). Choosing some observables as input parameters, such as the fine structure constant $\alpha$, the $Z$ boson mass $M_Z$, and the Fermi coupling constant $G_F$, other observables including the $W$ boson mass $M_W$ can be precisely predicted in the SM with radiative corrections. Any deviation of these measured observables from the SM prediction may provide a hint of NP.

Recently, the CDF collaboration reported an updated result of the direct measurement of the $W$ boson mass~\cite{CDF:2022hxs},
\begin{equation}
M_W =  80.4335\pm 0.0064_\mathrm{stat}\pm 0.0069_\mathrm{syst}~\si{GeV} = 80.4335\pm 0.0094~\si{GeV},
\end{equation}
using $8.8~\si{fb^{-1}}$ of data at the proton-antiproton collider Tevatron. Compared with the previous measurements performed by D0~\cite{D0:2012kms}, ATLAS~\cite{ATLAS:2017rzl}, LHCb~\cite{LHCb:2021bjt}, and CDF itself~\cite{CDF:2012gpf}, and the world average $M_W^\mathrm{ave}=80.379\pm 0.0012~\si{GeV}$~\cite{ParticleDataGroup:2020ssz}, the new CDF $M_W$ result has a smaller uncertainty but a much higher central value. This result also shows a significant deviation $\sim 7\sigma$ from the SM prediction $M_W^\mathrm{SM} = 80.3545\pm 0.0057~\si{GeV}$ obtained by the EW global fit \cite{deBlas:2021wap}, while the results of other EW measurements are basically consistent with the SM prediction. In order to confirm the deviation of $M_W$, more precise measurements are required in the future.

If the deviation of $M_W$ given by CDF is true, then it would be an evidence of NP at the EW sector, providing a very rich phenomenology \cite{Fan:2022dck,Zhu:2022tpr,Lu:2022bgw,Athron:2022qpo,Yuan:2022cpw,Strumia:2022qkt,Yang:2022gvz,deBlas:2022hdk,Du:2022pbp,Tang:2022pxh,Cacciapaglia:2022xih,Blennow:2022yfm,Arias-Aragon:2022ats,Zhu:2022scj,Sakurai:2022hwh,Fan:2022yly,Liu:2022jdq,Lee:2022nqz,Cheng:2022jyi,Song:2022xts,Bagnaschi:2022whn,Paul:2022dds,Bahl:2022xzi,Asadi:2022xiy,DiLuzio:2022xns,Athron:2022isz,Gu:2022htv,Heckman:2022the,Babu:2022pdn,Heo:2022dey,Du:2022brr,Cheung:2022zsb,Crivellin:2022fdf,Endo:2022kiw,Biekotter:2022abc,Balkin:2022glu,Krasnikov:2022xsi,Ahn:2022xeq,Han:2022juu,Zheng:2022irz,Kawamura:2022uft,Peli:2022ybi,Ghoshal:2022vzo,Perez:2022uil,Kanemura:2022ahw,Mondal:2022xdy,Zhang:2022nnh,Borah:2022obi,Chowdhury:2022moc,Arcadi:2022dmt,Cirigliano:2022qdm,Carpenter:2022oyg,Popov:2022ldh,Ghorbani:2022vtv,Du:2022fqv,Bhaskar:2022vgk,Batra:2022org,Cao:2022mif,Zeng:2022lkk,Baek:2022agi,Borah:2022zim,Almeida:2022lcs,Cheng:2022aau,Heeck:2022fvl,Addazi:2022fbj,Lee:2022gyf,Cai:2022cti,Benbrik:2022dja,Yang:2022qgs,Batra:2022pej,Tan:2022bip,Abouabid:2022lpg,Gisbert:2022lao,Chen:2022ocr,Zhou:2022cql,Gupta:2022lrt}. The universal framework of the effective field theory is helpful for understanding the properties of the corresponding NP. The general NP effects on the EW observables can be described by some well-known parameters, such as EW oblique parameters $S$, $T$, and $U$~\cite{Peskin:1990zt, Peskin:1991sw}. Through the global fit including the contributions of $\SUtwoL$ invariant dimension-6 operators, some groups have pointed out that the NP effects generating $|H^\dagger D_\mu H|^2$ with $T \gtrsim 0.1$ can easily explain the CDF $M_W$ anomaly \cite{Lu:2022bgw, deBlas:2022hdk, Strumia:2022qkt, Bagnaschi:2022whn, Fan:2022yly, Gu:2022htv, Balkin:2022glu}. Such interpretation can be realized by either the tree-level NP contributions at a few TeV or the loop-level NP contributions at a few hundred GeV in many specific models.

It is obvious that the introduction of new EW multiplets could affect the EW precision observables through loop corrections (e.g., Refs.~\cite{DEramo:2007anh, Dedes:2014hga, Fedderke:2015txa, Cai:2016sjz, Cai:2017wdu}). The lightest neutral NP particle in these multiplets could be a kind of so-called weakly interacting massive particles, serving as a candidate of cold dark matter (DM) under a $Z_2$ symmetry. On the other hand, the NP particles in the extra multiplets may also couple to the Higgs boson and affect its properties, such as the Higgs production and decay rates at colliders \cite{McCullough:2013rea, Freitas:2015hsa, Xiang:2017yfs} and the vacuum stability \cite{Elias-Miro:2012eoi, Chakrabarty:2014aya, DuttaBanik:2018emv, Khan:2016sxm, Wang:2018lhk}. Therefore, such models would provide a very rich phenomenology at both the particle physics and astrophysical detection~\cite{Mahbubani:2005pt, Cirelli:2005uq, Enberg:2007rp, Cohen:2011ec, Cheung:2013dua, Calibbi:2015nha, Cai:2015kpa, Tait:2016qbg, Horiuchi:2016tqw, Banerjee:2016hsk, Abe:2017glm, Wang:2017sxx, Cai:2017fmr, LopezHonorez:2017zrd, Abe:2019wku, Zeng:2019tlw, Liu:2020dok, Fraser:2020dpy, Gao:2021jip}.

In this work, we study the CDF $M_W$ anomaly in two DM models involving new $\mathrm{SU}(2)_\mathrm{L}$ multiplets, namely the singlet-triplet scalar DM (STSDM)~\cite{Cai:2017wdu} and singlet-doublet fermionic DM (SDFDM) ~\cite{Mahbubani:2005pt, DEramo:2007anh, Enberg:2007rp} models. In the STSDM model, an inert real scalar singlet and an inert complex scalar triplet are introduced. We perform a random scan in the 8-dimensional parameter space, and find the parameter points accounting for the CDF $M_W$ anomaly and satisfying the DM relic density observation and direct detection constraints.

The SDFDM model involves a Weyl singlet and two Weyl doublets. We perform a scan in the 4-dimensional parameter space of this model. However, no parameter point simultaneously explaining the CDF $M_W$ anomaly and fulfilling the DM phenomenological requirements is found in the scan. We explore the reason for this result in detail and perform some checks to confirm this result. Generally speaking, compared with fermionic models, scalar models have more interactions between the NP particles and the Higgs boson. Therefore, scalar models have more free parameters to help satisfy all experimental requirements.

This paper is organized as follows. In Sec.~\ref{sec:STSDM_model} we introduce the STSDM model and the framework for calculating the corrections to $M_W$. Then a random scan is carried out to find the parameter points simultaneously interpreting the CDF $M_W$ anomaly and satisfying the DM requirements. In Sec.~\ref{sec:SDFDM_model}, we introduce the SDFDM model, and discuss why there do not exist parameter points satisfying both the requirements of the $M_W$ anomaly and DM phenomenology. Finally, the discussions and conclusions are given in Sec.~\ref{sec:conclusions}.

\section{The STSDM model}
\label{sec:STSDM_model}

\subsection{Model}
\label{sec:STSDM_model_model}

The STSDM model~\cite{Cai:2017wdu} contains an inert real scalar singlet $S$ and an inert complex scalar triplet $\Delta$.
They do not carry hypercharge, living in the following $\SUtwoL \times \UoneY$ representations,
\begin{equation}
S \in (\mathbf{1}, 0),\quad
\Delta = \begin{pmatrix}
    \Delta^+\\
    \Delta^0\\
    \Delta^-\\
    \end{pmatrix} \in (\mathbf{3}, 0).
\end{equation}
Here $S$ and $\Delta^0$ are electrically neutral, while $\Delta^+$ and $\Delta^-$ carry electric charges $Q = +1$ and $Q = -1$, respectively.
%``Complex'' means that $\Delta$ is not self-conjugated, i.e., $(\Delta^+)^\dag \neq \Delta^-$ and $(\Delta^0)^\dag \neq \Delta^0$.
The neutral component of $\Delta$ can be decomposed as $\Delta^0 = (\phi + i a)/\sqrt{2}$, where $\phi$ and $a$ are real scalars.
The triplet can be described by a traceless type-$(1,1)$ $\SUtwoL$ tensor $\Delta^i_j$, whose components are $\Delta^1_2 = \Delta^+$, $\Delta^2_1 = \Delta^-$, and $\Delta^1_1 = - \Delta^2_2 = -\Delta^0 / \sqrt{2}$.
The Hermitian conjugate of $\Delta^i_j$ is denoted as $(\Delta^\dag)^j_i \equiv (\Delta^i_j)^\dag$, whose components are $(\Delta^\dag)^2_1 = (\Delta^+)^\dag$, $(\Delta^\dag)^1_2= (\Delta^-)^\dag$, and $(\Delta^\dag)^1_1 = - (\Delta^\dag)^2_2 = -(\Delta^0)^\dag / \sqrt{2}$. 

In order to make sure $S$ and $\Delta$ inert, we assume a $Z_2$ symmetry of $S \to -S$ and $\Delta \to -\Delta$, which remains unbroken after the breaking of EW gauge symmetry.
The gauge-invariant Lagrangian for the NP sector is
\begin{equation}
\mathcal{L}_\mathrm{NP} = \frac{1}{2}(\partial _\mu S)\partial ^\mu S + [(D_\mu \Delta )^\dag]^j_i D^\mu \Delta^i_j  - V(S, \Delta),
\end{equation}
where the scalar potential involving $S$ and $\Delta$ is given by
\begin{eqnarray}
V(S, \Delta) &=& \frac{1}{2} m_S^2{S^2} + m_\Delta ^2 (\Delta^\dag)^j_i \Delta^i_j + \frac{1}{2}(\mu _\Delta ^2\Delta _j^i\Delta _i^j + \mathrm{H.c.}) + \frac{1}{2}\lambda _{Sh} S^2 H^\dag_i H^i + \lambda _1 H_i^\dag \Delta _j^i(\Delta ^\dag)_k^j H^k
\nonumber\\
&& + \lambda_2 H_i^\dag (\Delta ^\dag)_j^i\Delta _k^j H^k - ( \lambda _3 H_i^\dag \Delta _j^i \Delta _k^j H^k + \lambda _4 SH_i^\dag \Delta _j^i H^j + \mathrm{H.c.}) + (\text{irrelevant terms}). \qquad
\end{eqnarray}
$H^i$ denotes the SM Higgs doublet, and $H_i^\dag = (H^i)^\dag$. The superscripts and subscripts of $\mathrm{SU}(2)$ tensors are related by 2-dimensional Levi-Civita symbols $\epsilon^{ij}$ and $\epsilon_{ij}$, e.g., $H^i=\epsilon^{ij} H_j$ and $H_i=\epsilon_{ij} H^j$. 
The omitted terms in $V(S, \Delta)$ are interaction terms among the $S$ and $\Delta$ fields, which are irrelevant to the discussions in this work.
The EW gauge interaction terms from the covariant derivatives can be found in Ref.~\cite{Cai:2017wdu}.

Note that $(\Delta^\dag)^j_i \Delta^i_j H^\dag_k H^k$ can be expressed as a linear combination of $H_i^\dag \Delta _j^i(\Delta ^\dag)_k^j H^k$ and $H_i^\dag (\Delta ^\dag)_j^i\Delta _k^j H^k$, while $\Delta^i_j \Delta^j_i H^\dag_k H^k = 2 H_i^\dag \Delta _j^i \Delta _k^j H^k$.
Thus, we have listed all independent, renormalizable terms in $V(S, \Delta)$ above.
We further assume $CP$ conservation in the scalar sector, and all the parameters in $V(S, \Delta)$ are real.

After the Higgs field $H^i(x)$ gains a nonzero vacuum expectation value $v$, the interaction terms among the scalar fields contribute to the mass terms of $S$ and $\Delta$.
In the unitary gauge, we have $H^1(x) = 0$ and $H^2(x) = [v+h(x)]/\sqrt{2}$.
The resulting mass terms of the NP scalars are given by
\begin{equation}\setlength{\arraycolsep}{.5em}
\mathcal{L}_\mathrm{M} =  - \frac{1}{2} m_a^2 a^2 - \frac{1}{2}\begin{pmatrix}
   S & \phi   \\
  \end{pmatrix} M_0^2 \begin{pmatrix}
   S  \\
   \phi   \\
 \end{pmatrix}  - \begin{pmatrix}
   (\Delta ^ + )^\dag & \Delta ^ -   \\
 \end{pmatrix} M_{\mathrm{C}}^2 \begin{pmatrix}
   \Delta ^ +   \\
   (\Delta ^ - )^\dag  \\
 \end{pmatrix}, 
\end{equation}
where $m_a^2 = m_\Delta ^2 - \mu _\Delta ^2 + (\lambda_+ + 2\lambda _3)v^2/4$, and the mass-squared matrices are
\begin{eqnarray}
M_0^2 &=&\setlength{\arraycolsep}{.5em} \begin{pmatrix}
   m_S^2 + \lambda _{Sh} v^2/2 & - \lambda _4 v^2/2  \\
   - \lambda _4 v^2/2 & m_\Delta ^2 + \mu_\Delta ^2 + (\lambda _+- 2\lambda _3)v^2/4  \\
\end{pmatrix},
\\
M_{\mathrm{C}}^2 &=&\setlength{\arraycolsep}{.5em} \begin{pmatrix}
m_\Delta ^2 + (\lambda_+ - \lambda_-) v^2/4 & \mu _\Delta ^2 - \lambda _3 v^2/2  \\
   \mu _\Delta ^2 - \lambda _3 v^2/2 & m_\Delta ^2 + (\lambda_+ + \lambda_-) v^2/4  \\
\end{pmatrix}.
\end{eqnarray}
Here we have defined $\lambda_\pm \equiv \lambda_1 \pm \lambda_2$.

$M_0^2$ and $M_{\mathrm{C}}^2$ can be diagonalized by orthogonal matrices $O_0$ and $O_\mathrm{C}$, parametrized as
\begin{equation}\setlength{\arraycolsep}{.5em}
O_0 = \begin{pmatrix}
   c_\beta  &  - s_\beta \\
   s_\beta  & c_\beta   \\
\end{pmatrix},\quad
O_\mathrm{C} = \begin{pmatrix}
   c_\theta  &  - s_\theta  \\
   s_\theta  & c_\theta   \\
\end{pmatrix},
\end{equation}
where the shorthand notations $s_\beta \equiv \sin\beta$ and $c_\beta \equiv \cos\beta$ are used.
Then we have
\begin{equation}\setlength{\arraycolsep}{.5em}
O_0^{\mathrm{T}} M_0^2 O_0 = \begin{pmatrix}
   \mu_1^2 &   \\
    & \mu_2^2  \\
\end{pmatrix},\quad
O_\mathrm{C}^{\mathrm{T}} M_{\mathrm{C}}^2 O_\mathrm{C} = \begin{pmatrix}
   m_1^2 &   \\
    & m_2^2  \\
\end{pmatrix},
\end{equation}
with
\begin{eqnarray}
\mu^2_{1,2} & = & \frac{1}{2} \left[ m_S^2 + m_\Delta^2 + \mu_\Delta^2 + \frac{1}{4}(2\lambda_{Sh} + \lambda_+ - 2\lambda_3)v^2 \mp \sqrt{r} \right],
\\
r &\equiv& \left[ m_S^2 - m_\Delta^2 - \mu_\Delta^2 - \frac{1}{4}(\lambda_+ - 2\lambda_3 - 2\lambda_{Sh})v^2\right]^2 + \lambda_4^2 v^4,
\\
m^2_{1,2} &=& m_\Delta^2 + \frac{v^2}{4} \left[\lambda_+ \mp \sqrt{\lambda_-^2 + 4\left(\frac{2\mu_\Delta^2}{v^2} - \lambda_3\right)^2}\right].
\end{eqnarray}
The mass hierarchy $\mu_1 \leq \mu_2$ and $m_1 \leq m_2$ is adopted.
The rotation angles $\beta$ and $\theta$ are determined by
\begin{equation}
\sin \beta  = \frac{ - (M_\mathrm{0}^2)_{12}}{\sqrt {(M_\mathrm{0}^2)_{12}^2 + [(M_\mathrm{0}^2)_{11} - \mu_2^2]^2} },\quad
\sin \theta = \frac{ - (M_\mathrm{C}^2)_{12}}{\sqrt {(M_\mathrm{C}^2)_{12}^2 + [(M_\mathrm{C}^2)_{11} - m_2^2]^2} }.
\end{equation}
The neutral mass eigenstates $X_i$ and the charged mass eigenstates $\Delta^+_i$ are defined by
\begin{equation}
\begin{pmatrix}
   S  \\
   \phi   \\
\end{pmatrix} = O_0\begin{pmatrix}
   X_1  \\
   X_2  \\
\end{pmatrix},
\quad
\begin{pmatrix}
   \Delta ^ +   \\
   (\Delta ^ - )^\dag  \\
\end{pmatrix} = O_\mathrm{C}\begin{pmatrix}
   \Delta _1^ +  \\
   \Delta _2^ +   \\
\end{pmatrix}.
\end{equation}
Expressing with the mass eigenstates, the mass terms become
\begin{equation}
\mathcal{L}_\mathrm{M} = 
 - \frac{1}{2}m_a^2 a^2 - \frac{1}{2}\sum\limits_{i = 1}^2 \mu _i^2X_i^2  - \sum\limits_{i = 1}^2 m_i^2\Delta _i^ - \Delta _i^ + ,
\end{equation}
where $\Delta _i^ - \equiv (\Delta _i^ +)^\dag$.

Now we have one $CP$-odd neutral scalar boson $a$, two $CP$-even neutral scalar bosons $X_1$ and $X_2$, and two singly charged scalar bosons $\Delta^+_1$ and  $\Delta^+_2$ accompanied with the antiparticles.
They are all $Z_2$-odd, and the lightest one is stable, ensured by the $Z_2$ symmetry.
There are eight free parameters in the model, $m_S^2$, $m_\Delta ^2$, $\mu _\Delta ^2$, $\lambda_{Sh}$, $\lambda_+$, $\lambda_-$, $\lambda_3$, and $\lambda_4$.
It can be proved that $m_a \geq m_1$.
Only if $\lambda_- = 0$ and $2\mu _\Delta ^2/v^2 - \lambda _3 \geq 0$, we have $m_a = m_1$, otherwise $a$ is heavier than $\Delta^\pm_1$ and cannot be a DM candidate.
Hereafter we assume that $X_1$ is the lightest NP scalar, i.e., $\mu_1 < m_a$ and $\mu_1 < m_1$.
Thus, the DM candidate in this model is $X_1$, whose mass is $\mu_1$.

The NP corrections to $M_W$ can be given in terms of the EW oblique parameters $S$, $T$, and $U$ as~\cite{Ciuchini:2013pca}
\begin{equation}
\Delta M_W = -\frac{\alpha M_W^{\rm SM}}{4(\cw-\sw)} \left(S-2\cw T- \frac{\cw-\sw}{2 \sw} U \right) \simeq  0.44 \; (T - 0.64\, S + 0.80\, U)\; \rm{GeV},
\end{equation}
where $s_\mathrm{W} \equiv \sin \theta_{\rm W}$, $c_\mathrm{W} \equiv \cos \theta_{\rm W}$, and $\theta_{\rm W}$ is the Weinberg angle.
$M_W^{\rm SM}$ is the SM prediction of the $W$ boson mass including radiation corrections. The SM corresponds to the case with $S = T = U = 0$.
Denoting $\Pi_{ij}(p^2)$ to be the NP contributions to the $g_{\mu\nu}$ coefficients of the vacuum polarizations of EW gauge bosons $i$ and $j$, the oblique parameters are defined by~\cite{Peskin:1990zt, Peskin:1991sw}
\begin{eqnarray}
S&\equiv&
16\pi[\Pi'_{33}(0) - \Pi'_{3Q}(0)]
= \frac{16\pi s_\mathrm{W}^2 c_\mathrm{W}^2}{e^2}\left[\Pi'_{ZZ}(0)-
\frac{c_\mathrm{W}^2 - s_\mathrm{W}^2}{s_\mathrm{W} c_\mathrm{W}} \Pi'_{ZA}(0)-\Pi'_{AA}(0)\right],\\
T&\equiv&
\frac{4\pi}{\sw \cw M_Z^2}[\Pi_{11}(0) - \Pi_{33}(0)]
= \frac{4\pi}{e^2}\left[\frac{\Pi_{WW}(0)}{M_W^2}-\frac{\Pi_{ZZ}(0)}{M_Z^2}\right],\\
U&\equiv&
16\pi[\Pi'_{11}(0) - \Pi'_{33}(0)]
= \frac{16\pi s_\mathrm{W}^2}{e^2}\left[\Pi'_{WW}(0)-c_\mathrm{W}^2\Pi'_{ZZ}(0)
-2 s_\mathrm{W} c_\mathrm{W} \Pi'_{ZA}(0)
-s_\mathrm{W}^2 \Pi'_{AA}(0)\right],\qquad
\end{eqnarray}
where $\Pi'_{ij}(0) \equiv \partial \Pi_{ij}(p^2)/\partial p^2 |_{p^2=0}$ and $e$ is the electric charge.

Because of EW gauge interactions and mass mixings from spontaneous symmetry breaking, the NP scalars contribute to the self-energies of EW gauge bosons, leading to~\cite{Cai:2017wdu}
\begin{eqnarray}
\Pi_{11}(p^2) &=& \frac{1}{16\pi ^2}\big\{ 2(1 - s_{2\theta })B_{00}(p^2,m_a^2,m_1^2) + 2(1 + s_{2\theta })B_{00}(p^2,m_a^2,m_2^2)
\nonumber\\
&&\hspace{3em} + 2(1 + s_{2\theta })[s_\beta ^2 B_{00}(p^2,\mu _1^2,m_1^2) + c_\beta ^2 B_{00}(p^2,\mu _2^2,m_1^2)]
\nonumber\\
&&\hspace{3em} + 2(1 - s_{2\theta })[s_\beta ^2 B_{00}(p^2,\mu _1^2,m_2^2) + c_\beta ^2 B_{00}(p^2,\mu _2^2,m_2^2)]
\nonumber\\
&&\hspace{3em} - [A_0(m_1^2) + A_0(m_2^2) + A_0(m_a^2) + s_\beta ^2 A_0(\mu _1^2) + c_\beta ^2 A_0(\mu _2^2)]\big\},
\\
\Pi _{3Q}(p^2) &=& \Pi _{33}(p^2) = \frac{1}{16\pi ^2}[4B_{00}(p^2,m_1^2,m_1^2) + 4B_{00}(p^2,m_2^2,m_2^2) - 2A_0(m_1^2) - 2A_0(m_2^2)],\quad
\end{eqnarray}
where $A_0$ and $B_{00}$ are the  standard Passiano-Veltman scalar functions~\cite{Passarino:1978jh}.
From these expressions, one can compute the EW oblique parameters.
Because of $2B_{00}(0,m^2,m^2) = A_0(m^2)$, we have $\Pi _{3Q}(0) = \Pi _{33}(0) = 0$ and hence $T = 4\pi \Pi_{11}(0)/(\sw \cw M_Z^2)$.
The $S$ parameter is related to the hypercharge gauge field by definition.
Since neither the singlet nor the triplet carries hypercharge, $\Pi _{3Q}(p^2) = \Pi _{33}(p^2)$ always holds, resulting in $S = 0$ in this model.
On the other hand, the $T$ and $U$ parameters would vanish when the custodial symmetry is respected~\cite{Sikivie:1980hm, Peskin:1991sw}.
The condition for the custodial symmetry in the model is $\lambda_- = \lambda_4 = 0$~\cite{Cai:2017wdu}.

Expressing the trilinear interaction term between $X_1$ and the Higgs boson $h$ as $\mathcal{L}_{hX_1^2} =  \lambda _{hX_1^2} v h X_1^2/2$, the dimensionless $h X_1^2$ coupling is
\begin{equation}
\lambda _{h X_1^2} =  - \lambda _{Sh}c_\beta ^2 - \frac{1}{2}(\lambda _ +  - 2\lambda _3)s_\beta ^2 + 2\lambda _4 s_\beta c_\beta .
\end{equation}
This coupling induces a spin-independent (SI) $X_1$-nucleon scattering cross section of~\cite{Yu:2011by}
\begin{equation}
\sigma _N^{\mathrm{SI}} = \frac{m_N^2 F_N^2}{4\pi (\mu _1 + m_N)^2},
\end{equation}
with
\begin{equation}
F_N =  - \frac{\lambda_{h X_1^2} m_N}{9m_h^2}[2 + 7(f_u^N + f_d^N + f_s^N)],
\end{equation}
where $f_q^N$ are nucleon form factors.

\subsection{Results}
\label{sec:STSDM_model_results}

In this section, we perform a random scan in the 8-dimensional parameter space of the STSDM model within the following ranges:
\begin{equation}
(10~\si{GeV})^2 < |m_S^2|, |m_\Delta^2|, |\mu_\Delta^2| < (1~\si{TeV})^2,\quad
0.001 < |\lambda_{Sh}|, |\lambda_+|, |\lambda_-|, |\lambda_3|, |\lambda_4| < 1.
\end{equation}
The EW oblique parameters $T$ and $U$ predicted by each parameter point are calculated based on the above expressions in the help of the code \texttt{LoopTools}~\cite{Hahn:1998yk}.
We implement the model with \texttt{FeynRules}~\cite{Christensen:2009jx,Alloul:2013bka} and utilize the package \texttt{micrOMEGAs~5.2.13}~\cite{Belanger:2020gnr} to compute the $X_1$ relic density $\Omega_c h^2$, the SI $X_1$-nucleon scattering cross section $\sigma^\mathrm{SI}_N$, and the thermally averaged $X_1X_1$ annihilation cross section today $\left<\sigma v\right>$.

Firstly, we require the selected parameter points leading to moderate $T$ or $U$ parameters, that can explain the CDF $M_W$ anomaly and satisfy other EW precision tests.
Here we adopt the result of a global fit of the oblique parameters to EW precision observables including the recent CDF $M_W$ measurement~\cite{Lu:2022bgw}:
\begin{equation}
S = 0.06 \pm 0.10,\quad
T = 0.11 \pm 0.12,\quad
U = 0.14 \pm 0.09
\end{equation}
with correlation coefficients
\begin{equation}
\rho_{ST} = 0.90,\quad
\rho_{SU} = -0.59,\quad
\rho_{TU} = -0.85.
\end{equation}
Note that the fitted $S$, $T$, and $U$ are highly correlated, and the correlation coefficients are rather important when evaluating the deviation of the model prediction to the experimental results.
The predicted $T$ and $U$ in the STSDM model are required to be within the $95\%$ C.L. region of the fit result, based on a $\chi^2$ calculation.
Then we demand the predicted $X_1$ thermal relic density within the $3\sigma$ range of the Planck measurement $\Omega_c h^2=0.1200 \pm 0.0012$~\cite{Planck:2018vyg} and the SI $X_1$-nucleon scattering cross section satisfies the constraint from the PandaX-4T direct detection experiment at $90\%$ C.L.~\cite{PandaX-4T:2021bab}.

\begin{figure}[!t]
\centering
\subfigure[~$M_W$-$T$ plane \label{fig:STSDM_TU:mw}]{\includegraphics[width=0.48\textwidth]{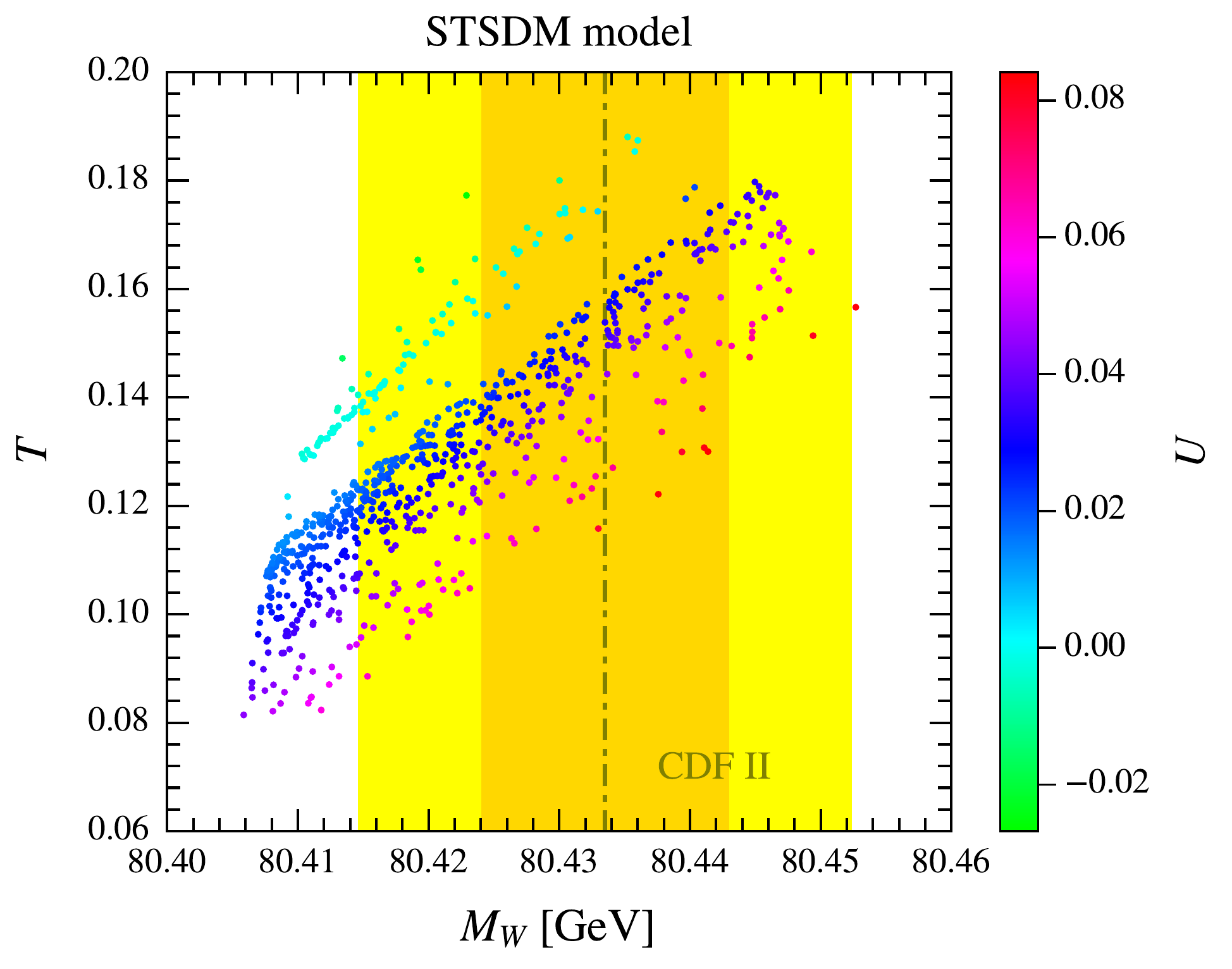}}
\subfigure[~$T$-$U$ plane\label{fig:STSDM_TU:mu1}]{\includegraphics[width=0.48\textwidth]{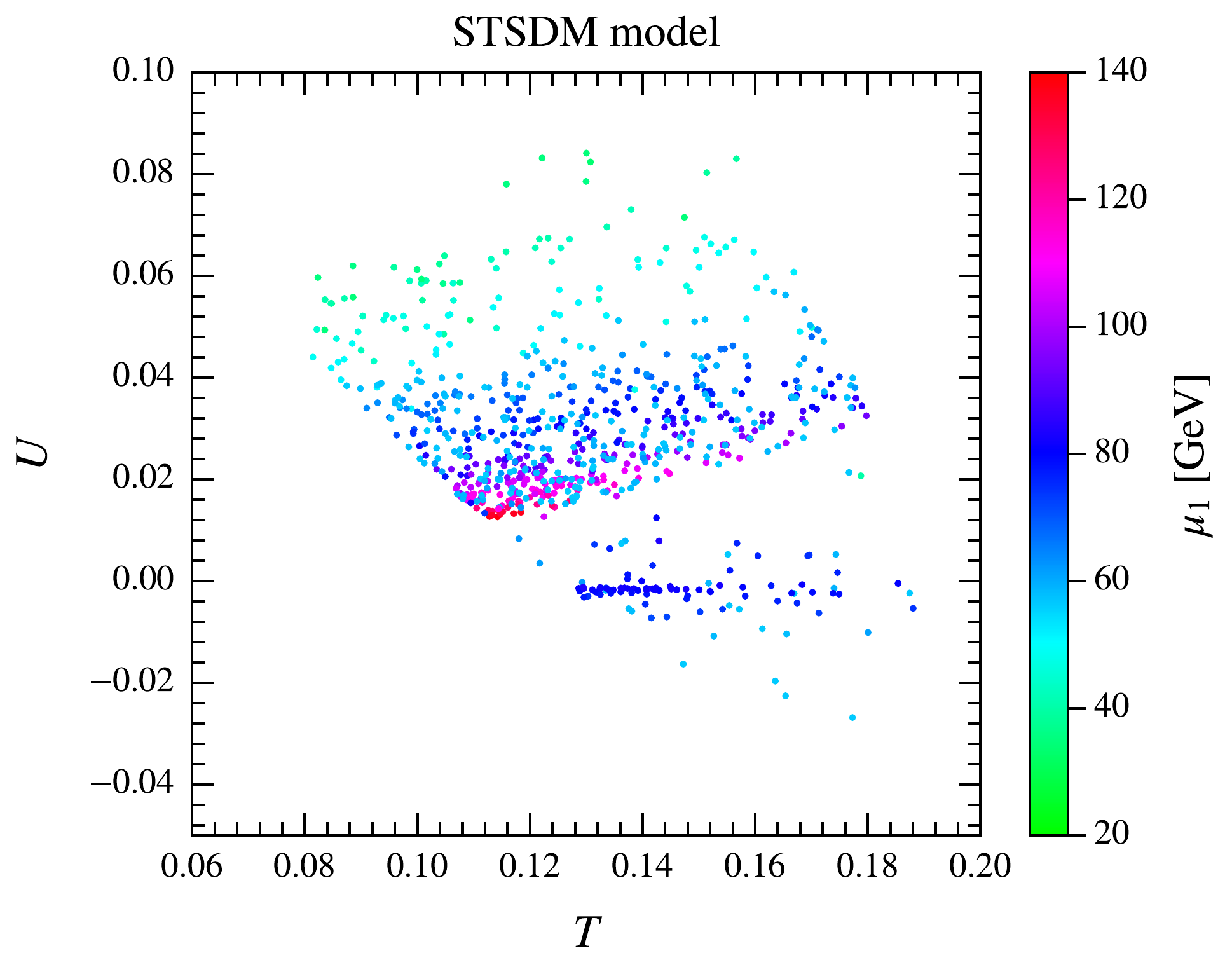}}
\caption{Selected parameter points projected in the $M_W$-$T$ (a) and $T$-$U$ (b) planes for the STSDM model, with the colors corresponding to $U$ and the DM mass $\mu_1$, respectively.
In the left panel, the dot-dashed line denotes the central value of the recent CDF $M_W$ measurement~\cite{CDF:2022hxs}, with gold and yellow colors indicating the $1\sigma$ and $2\sigma$ ranges, respectively.}
\label{fig:STSDM_TU}
\end{figure}

In Fig.~\ref{fig:STSDM_TU:mw}, the selected parameter points projected in the $M_W$-$T$ plane.
Most of the points lie within the $2\sigma$ range of the CDF $M_W$ result.
The other parameter points are out of the $2\sigma$ range but still consistent with the EW global fit at $95\%$ C.L.
Figure~\ref{fig:STSDM_TU:mu1} displays the parameter points in the $T$-$U$ plane with colors indicating the DM mass $\mu_1$.
The allowed region in the $T$-$U$ plane shows a negative correlation as the fit result of $\rho_{TU}$ implies.
A large fraction of the points correspond to $0.07 \lesssim T \lesssim 0.18$ and $0 \lesssim U \lesssim 0.05$.
In the framework of effective field theory, $T$ corresponds to a dimension-6 operator $|H^\dag D_\mu H|^2/\Lambda^2$, while $U$ corresponds to a dimension-8 operator $|H^\dag W^a_{\mu\nu} \sigma^a H|^2/\Lambda^4$.
A large cutoff scale $\Lambda$ would give more suppression on $U$ than $T$.
Nonetheless, the NP scalars in the loops here have masses around or even below the electroweak scale, so the effective operator description breaks down, and $U \sim T$ can be achieved.

A pure $\SUtwoL$ triplet scalar with EW interactions is expected to give the correct DM relic density for masses around $2~\si{TeV}$~\cite{Cirelli:2005uq}.
For interpreting the CDF $M_W$ anomaly, however, the DM candidate $X_1$ is required to be lighter than $150~\si{GeV}$, as shown in Fig.~\ref{fig:STSDM_TU:mu1}, and $X_1$ should have a moderate triplet component to contribute to $T$ and $U$.
Current direct detection experiments have strictly constrained the $h X_1 X_1$ coupling $\lambda _{h X_1^2}$ in such a mass range, typically leading to overproduction of the DM relic due to insufficient $X_1 X_1$ annihilation.
Nonetheless, it is well known that there are some exceptions for deriving the correct relic density with some special NP mass spectra~\cite{Griest:1990kh}. If coannihilation, resonant annihilation, or annihilation into forbidden channels has a significant contribution at the freeze-out epoch, then the observed relic density could be achieved for $\mu_1 < 150~\si{GeV}$.

\begin{figure}[!t]
\centering
\subfigure[~$\mu_1$-$m_1$ plane\label{fig:STSDM_mass:mu1_m1}]{\includegraphics[height=0.41\textwidth]{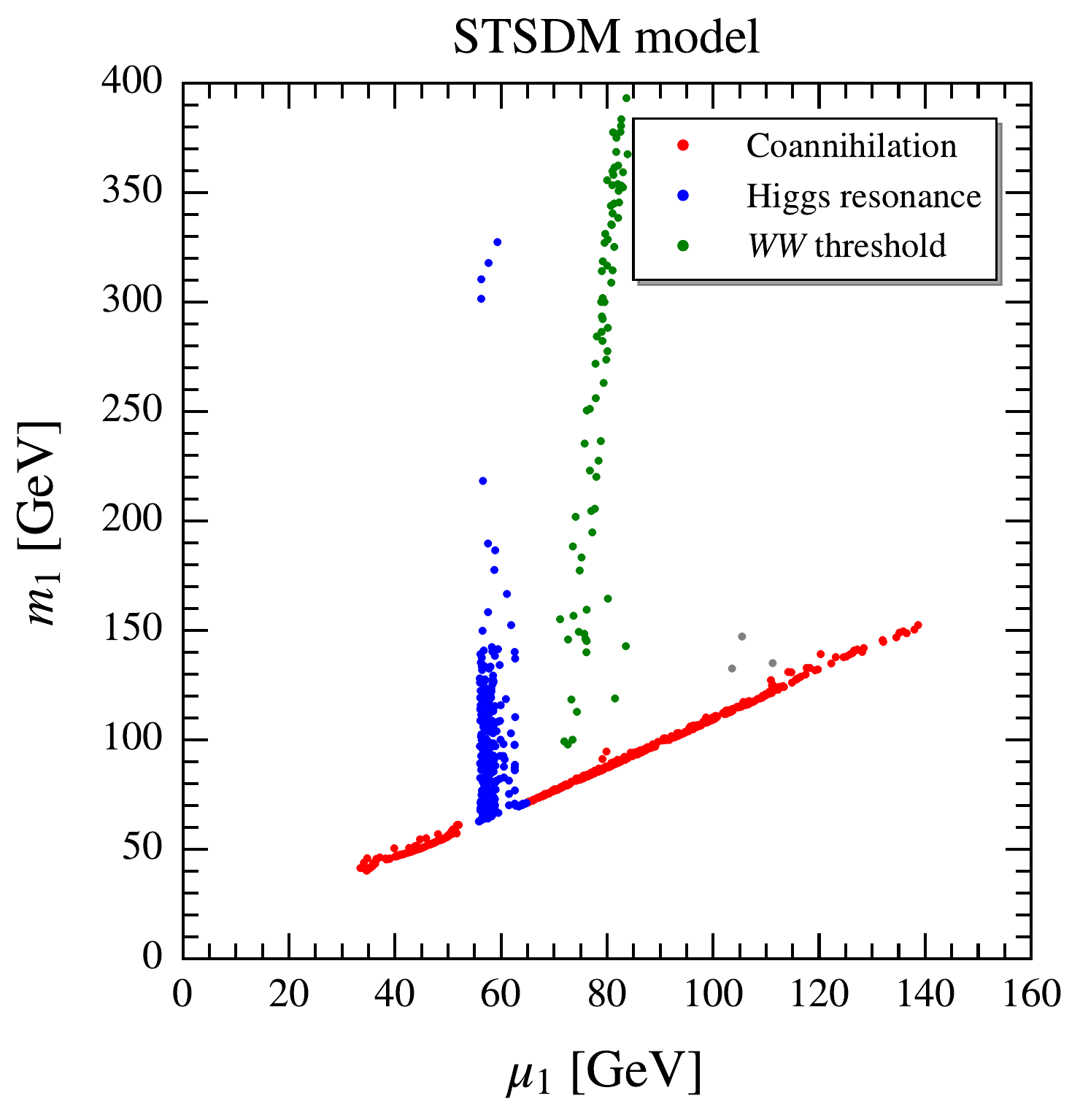}}
\hspace{1em}
\subfigure[~$\mu_2$-$m_a$ plane\label{fig:STSDM_mass:mu2_ma}]{\includegraphics[height=0.41\textwidth]{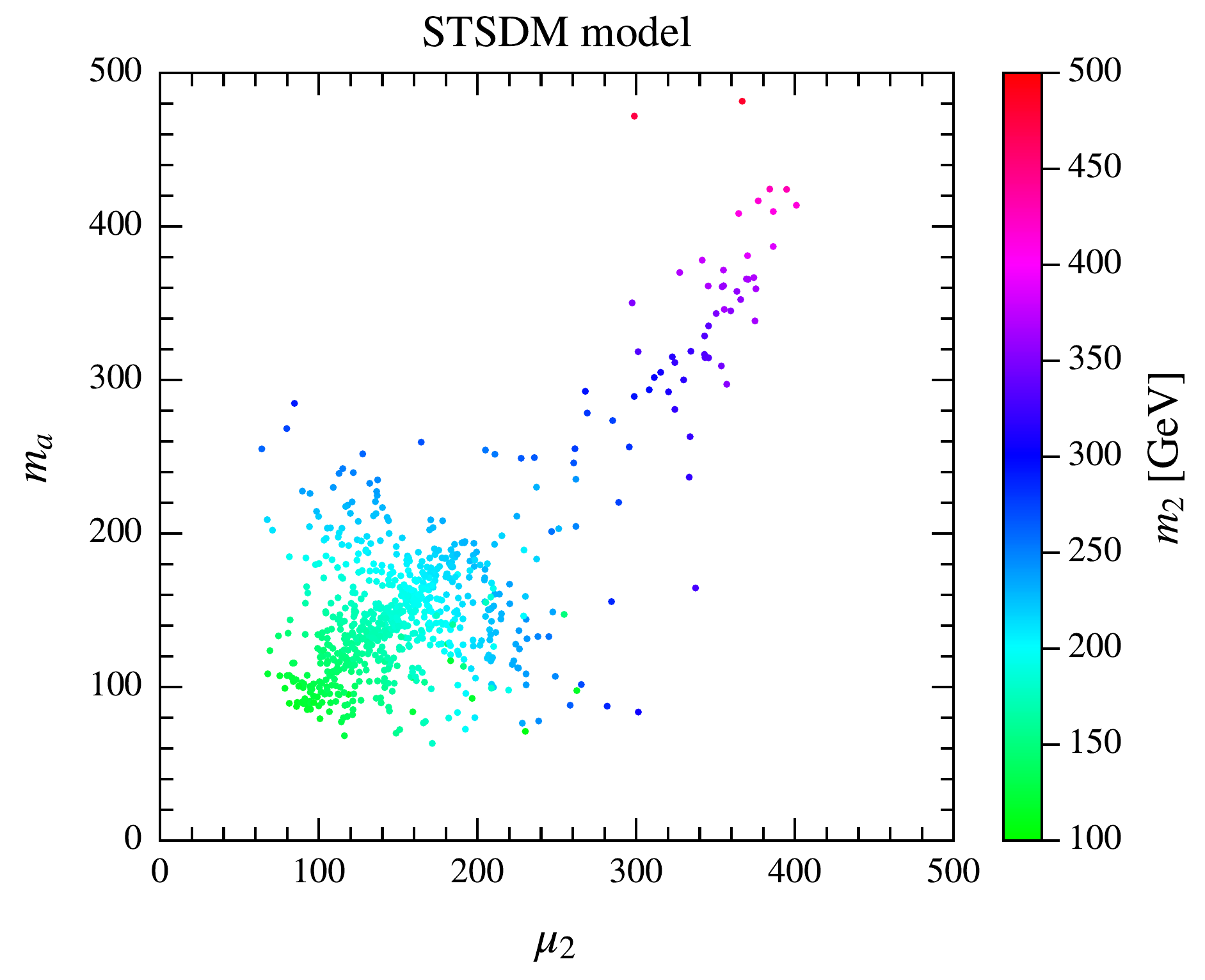}}
\caption{Selected parameter points projected in the $\mu_1$-$m_1$ (a) and $\mu_2$-$m_a$ (b) planes for the STSDM model.
The colors in the left panel indicate three exceptional categories for achieving the correct relic density.
The colors in the right panel denotes the $\Delta^\pm_2$ mass $m_2$.}
\label{fig:STSDM_mass}
\end{figure}

In Fig.~\ref{fig:STSDM_mass:mu1_m1}, we project the selected parameter points into the $\mu_1$-$m_1$ plane, with colors denoting three exceptional categories of points.
Firstly, the red points align around the $m_1 = \mu_1$ line would have an important $\Delta_1^\pm$-$X_1$ coannihilation effect at the freeze-out epoch, because $\Delta_1^\pm$ is just a little heavier than $X_1$ and could effectively annihilate into SM particles.
Secondly, the blue points with $\mu_1 \sim m_h/2$ lead to resonant annihilation of $X_1 X_1$ through a $s$-channel Higgs boson.
Thirdly, the green points with $\mu_1$ slightly smaller than $M_W$ are corresponding to the $WW$ threshold effect. In this case, the 2-body annihilation channel $X_1 X_1 \to W^+W^-$ is forbidden today, but opens at the freeze-out epoch due to higher center-of-mass energies at the temperature $\sim \mu_1/25$.
Figure~\ref{fig:STSDM_mass:mu2_ma} demonstrates the masses of the $a$, $X_2$, and $\Delta^\pm_2$ scalar bosons, which can be as large as $400\text{--}500~\si{GeV}$.
But a lot of parameter points correspond to the masses less than $200~\si{GeV}$.

\begin{figure}[!t]
\centering
\subfigure[~$\mu_1$-$\sigma^\mathrm{SI}_N$ plane\label{fig:STSDM_DD}]{\includegraphics[width=0.46\textwidth]{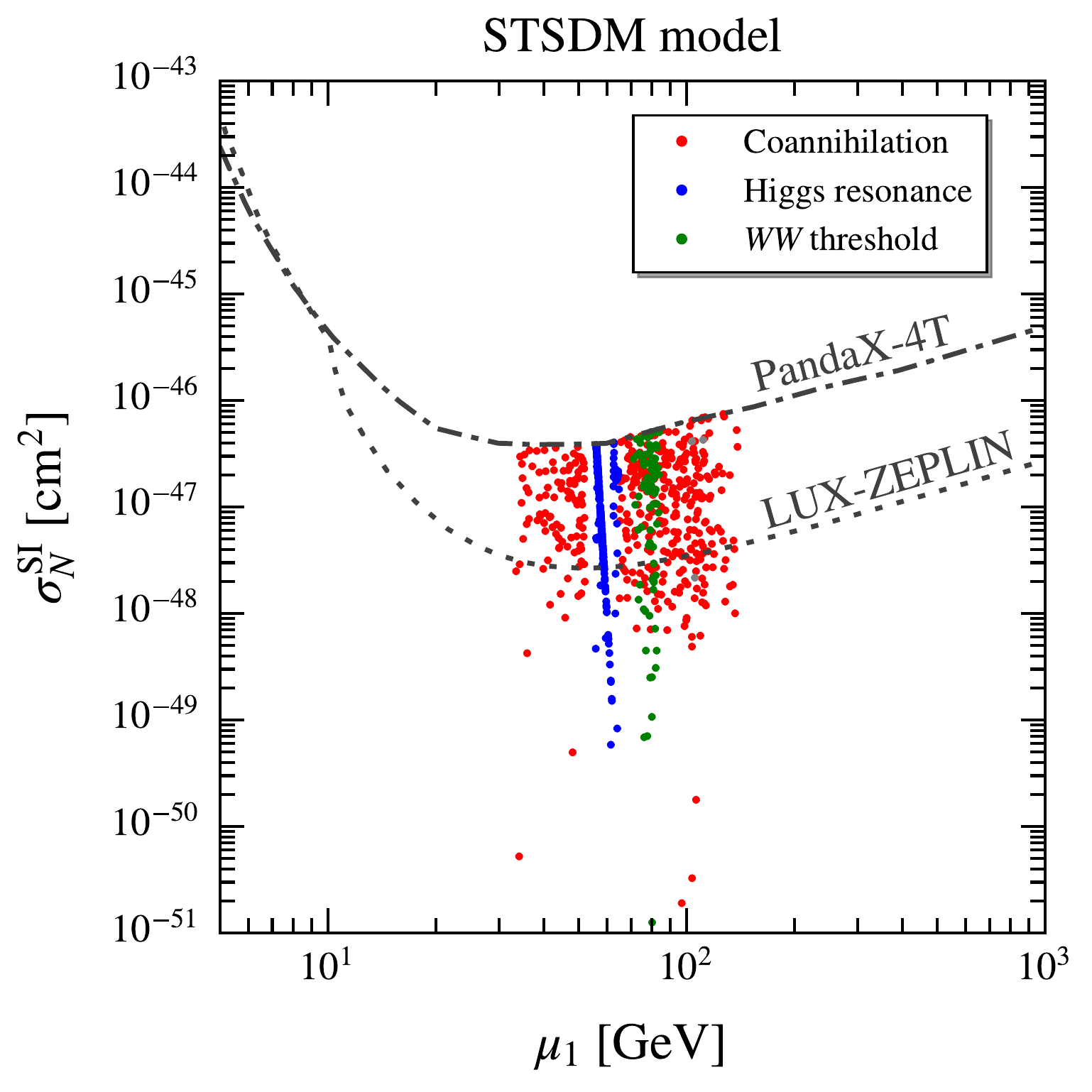}}
\hspace{1em}
\subfigure[~$\mu_1$-$\left<\sigma v\right>$ plane\label{fig:STSDM_ID}]{\includegraphics[width=0.46\textwidth]{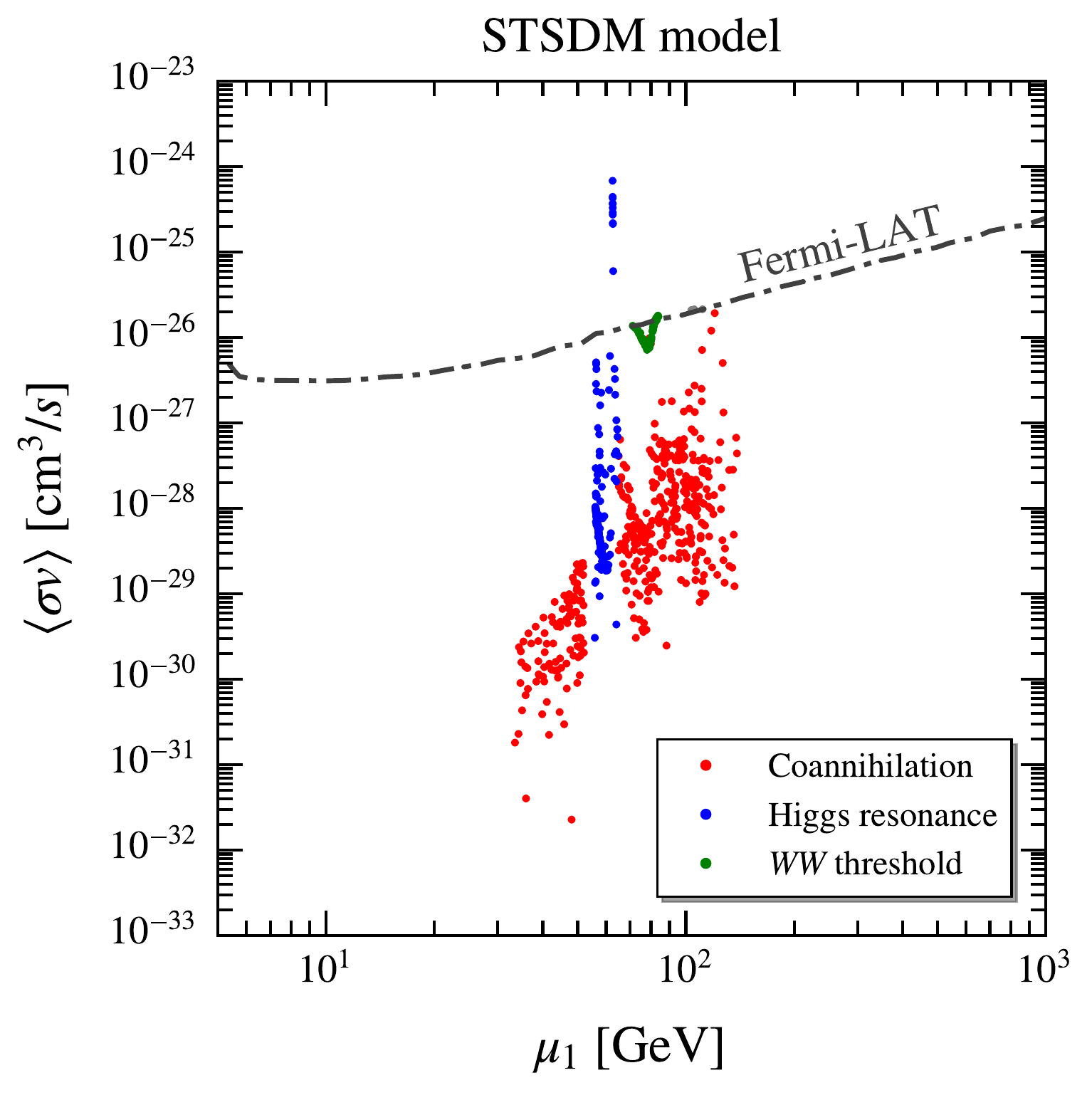}}
\caption{Selected parameter points projected in the $\mu_1$-$\sigma^\mathrm{SI}_N$ (a) and $\mu_1$-$\left<\sigma v\right>$ (b) planes for the STSDM model, with the colors indicating three exceptional categories for achieving the correct relic density.
In the left panel, the dot-dashed line denotes the $90\%$ C.L. upper limit from the PandaX-4T direct detection experiment~\cite{PandaX-4T:2021bab}, while the dotted line demonstrates the future sensitivity of the LUX-ZEPLIN experiment~\cite{Mount:2017qzi}.
The dot-dashed line in the right panel shows the $95\%$ C.L. upper limit for the $b\bar{b}$ annihilation channel from the Fermi-LAT $\gamma$-ray observations of dwarf galaxies~\cite{Hoof:2018hyn}.}
\label{fig:STSDM_DD_ID}
\end{figure}

The SI $X_1$-nucleon scattering cross sections corresponding to the selected points are shown in Fig.~\ref{fig:STSDM_DD}.
Some parameter points seem far from the reach of the current PandaX-4T direct detection experiment, but a large fraction of them could be properly tested in the future LUX-ZEPLIN experiment~\cite{Mount:2017qzi}.

Figure~\ref{fig:STSDM_ID} displays the total $X_1 X_1$ annihilation cross sections at rest as well as the constraint from the 11-year Fermi-LAT $\gamma$-ray observations of 27 dwarf spheroidal galaxies in the $b\bar{b}$ annihilation channel~\cite{Hoof:2018hyn}. Most of the points with $\left<\sigma v\right>$ lower than the standard value $\sim 10^{-26}~\si{cm^3/s}$ are safe. Note that for the parameter points with $\mu_1$ slightly smaller than $M_W$, although the 2-body annihilation channel $X_1 X_1 \to W^+W^-$ is forbidden today, the 3-body annihilation $X_1 X_1 \to W^+W^{-*}$ or $W^{+*}W^{-}$ could still contribute a moderate $\left<\sigma v\right>$.
A few parameter points with the resonant annihilation have been excluded by the Fermi-LAT search. This is because that for some parameter points with the $X_1X_1$ annihilation near a very narrow resonance, the annihilation cross section would increase with the decreasing  velocities of $X_1$. This enhancement could lead to a too large $\left<\sigma v\right>$ today.   

In order to explain the CDF $M_W$ anomaly, the masses of the NP scalar bosons are required to be smaller than a few hundred GeV. These NP particles could be directly searched at the past and current colliders.
If $\mu_1 < m_h/2$, the invisible decay channel $h \to X_1 X_1$ would open and affect the properties of the $125~\si{GeV}$ Higgs boson.
Therefore, we also consider bounds from collider experiments in the scan.

The selected parameter points demonstrated above have also passed the following three tests.
First, the signal strengths of the $125~\si{GeV}$ Higgs boson are required to be consistent with current LHC measurements at 95\% C.L., based on a global likilihood function constructed by \texttt{Lilith~2}~\cite{Kraml:2019sis} implemented in \texttt{micrOMEGAs}.
Second, we utilize \texttt{SModelS~2}~\cite{Alguero:2021dig} interfaced by \texttt{micrOMEGAs} to test the parameter points according to the exclusion limits from LHC direct searches for pair production of the NP scalars.
Third, we simulate the monojet signature induced by the STSDM model at the 13~TeV LHC with the Monte Carlo generator \texttt{MadGraph5\_aMC@NLO}~\cite{Alwall:2014hca}, and reinterpret the ATLAS monojet analysis for an integrated luminosity of $139~\si{fb^{-1}}$ at $\sqrt{s} = 13~\si{TeV}$~\cite{ATLAS:2021kxv} to set 95\% C.L. exclusion limits on the parameter points.
We find that the monojet search has excluded the DM candidate mass $\mu_1$ up to $\sim 33~\si{GeV}$.

In addition, the charged scalars $\Delta^\pm_{1,2}$ could be directly produced in pairs at LEP.
A similar search for the lightest chargino by the DELPHI collaboration has excluded the chargino mass up to $103.4~\si{GeV}$~\cite{DELPHI:2003uqw}.
Thus, we expect that the parameter points with $m_{1,2} \lesssim 100~\si{GeV}$ might be already excluded.
However, such a bound could be relieved if the mass splitting between $\Delta^\pm_1$ and $X_1$ is rather small.
Therefore, the constraint on the coannihilation category should be weaker, and a considerable number of the selected parameter points should be free from the LEP and LHC bounds.

\section{The SDFDM model}
\label{sec:SDFDM_model}

\subsection{Model}
\label{sec:SDFDM_model_model}

The SDFDM model contains a left-handed Weyl singlet $S$ and two left-handed Weyl doublets $D_1$ and $D_2$ with opposite hypercharges.
They live in the following $\SUtwoL \times \UoneY$ representations:
\begin{equation}
	S \in (\mathbf{1}, 0),\quad
    D_1 = \begin{pmatrix}
    D_1^0\\
    D_1^-
    \end{pmatrix} \in (\mathbf{2}, -1/2),\quad
    D_2 = \begin{pmatrix}
    D_2^+ \\
    D_2^0
    \end{pmatrix} \in (\mathbf{2}, 1/2).
\end{equation}
Here $D_1^0$ and $D_2^0$ are electrically neutral components, while $D_1^-$ and $D_2^+$ are singly charged components.
The gauge-invariant Lagrangian for the NP sector reads
\begin{eqnarray}
\mathcal{L}_\mathrm{NP} &=& i S^\dagger \bar{\sigma}^\mu \partial_\mu S - \frac{1}{2} (m_S S S + \mathrm{H.c.})
 + iD_1^\dag {{\bar \sigma }^\mu }{D_\mu }{D_1} + iD_2^\dag {{\bar \sigma }^\mu }{D_\mu }{D_2} - ({m_D} \epsilon_{ij} {D_1^i}{D_2^j} + \mathrm{H.c.})
\nonumber\\
&& + (y_1 SD_1^i{H_i} - y_2 SD_2^iH_i^\dag  + \mathrm{H.c.}),
\end{eqnarray}
where the covariant derivative for $D_1$ and $D_2$ is $D_\mu = \partial_\mu - ig W_\mu^a \sigma_a/2 - i g' Y B_\mu$.
Assuming $CP$ conservation in the NP sector, the model have four real parameters, including two mass parameters $m_S$ and $m_D$, and two Yukawa couplings $y_1$ and $y_2$. Without loss of generality, we allow $y_1$ and $y_2$ to be positive or negative, while fix $m_S, m_D>0$.

As the Higgs field develops a vacuum expectation value $v$, the Yukawa couplings induce Dirac mass terms for the NP fermions.
Consequently, the fermion mass terms are given by
\begin{equation}\setlength{\arraycolsep}{.5em}
\mathcal{L}_\mathrm{M} =
-\frac{1}{2}
\begin{pmatrix}
S & D_1^0 & D_2^0
\end{pmatrix}
M_\mathrm{N}
\begin{pmatrix}
S\\
D_1^0\\
D_2^0
\end{pmatrix}
- m_D D_1^- D_2^+ + \mathrm{H.c.}\,,
\end{equation}
where the mass matrix for the neutral fermions is
\begin{equation}\setlength{\arraycolsep}{.5em}
M_\mathrm{N} = 
\begin{pmatrix}
m_S & y_1 v/\sqrt{2} & y_2 v/\sqrt{2}\\
y_1 v/\sqrt{2} & 0 &- m_D \\
y_2 v/\sqrt{2} & -m_D & 0
\end{pmatrix}.
\end{equation}
This matrix can be diagonalized by an orthogonal matrix $N$, i.e., $N^\mathrm{T} M_\mathrm{N} N = \operatorname{diag} (m_{\chi^0_1}, m_{\chi^0_2}, m_{\chi^0_3})$.
Here we adopt a mass hierarchy of $m_{\chi^0_1} \leq m_{\chi^0_2} \leq m_{\chi^0_3}$.

Defining three neutral mass eigenstates $\chi^0_i$ by
\begin{equation}
\begin{pmatrix}
\chi_1^0\\
\chi_2^0\\
\chi_3^0
\end{pmatrix}
= N^\mathrm{T}
\begin{pmatrix}
S\\
D_1^0\\
D_2^0
\end{pmatrix},
\end{equation}
the mass terms are expressed as
\begin{equation}
\mathcal{L}_\mathrm{M} = -\frac{1}{2}\sum_{i=1}^3 m_{\chi_i^0} \chi_i^0\chi_i^0 - m_{\chi^\pm}\chi^-\chi^+ +\mathrm{H.c.}\,,
\end{equation}
with $m_{\chi^\pm} = m_D$, $\chi^+ = D_2^+$, and $\chi^- = D_1^-$.
Now we have three Majorana fermions $\chi^0_i$ and two singly charged fermions $\chi^\pm$.
Assuming a $Z_2$ symmetry, the lightest Majorana fermions$\chi^0_1$ can act as a DM candidate. The full Lagrangian describing the interactions between these NP fermions and the SM particles, including the EW gauge bosons and the Higgs boson, can be found in Refs.~\cite{Cai:2016sjz,Xiang:2017yfs}. These interactions would affect $M_W$ and other EW precision observables via loop corrections.
The detailed expressions of $\Pi_{WW}(p^2)$, $\Pi_{ZZ}(p^2)$, $\Pi_{ZA}(p^2)$, and $\Pi_{AA}(p^2)$ contributed by the NP fermions in the SDFDM model are provided in Ref.~\cite{Cai:2016sjz}. Once the fermion masses and couplings to the EW gauge bosons are given as inputs, we can obtain the numerical values of the oblique parameters.

The current results of DM direct detection set stringent constraints on the parameter space of the SDFDM model~\cite{Cai:2016sjz,Xiang:2017yfs}. The Yukawa couplings give rise to the coupling $g_{h\chi_1^0\chi_1^0}$ between $\chi_1^0$ and the Higgs boson $h$. This coupling leads to the SI DM-nucleon scattering with a cross section of~\cite{Zheng:2010js}
\begin{equation}
\sigma_N^\mathrm{SI}=\frac{4\mu_{\chi N}^2 G_{\mathrm{S},N}^2}{\pi},
\end{equation}
where $\mu_{\chi N}\equiv m_{\chi_1^0}m_N/(m_{\chi_1^0}+m_N)$ is the DM-nucleon reduced mass, and
\begin{equation}
G_{\mathrm{S},N}=\frac{N_{11} m_N}{9\sqrt{2}v M_h^2}(y_1N_{21}+y_2N_{31})[2 + 7 (f^N_u + f^N_d + f^N_s)]
\end{equation}
is the scalar-type effective coupling.
The $\chi^0_1$ coupling to the $Z$ boson $g_{Z\chi_1^0\chi_1^0}$ from the EW gauge interaction also induces a spin-dependent (SD) DM-nucleon scattering cross section~\cite{Zheng:2010js}
\begin{equation}
\sigma_N^\mathrm{SD}=\frac{12}{\pi}\mu_{\chi N}^2 G_{\mathrm{A},N}^2,
\end{equation}
where the axial-vector effective coupling is
\begin{equation}
G_{\mathrm{A},N}=\sum_{q=u,d,s}\frac{g^q_\mathrm{A} g^2 \Delta_q^N}{8 c_\mathrm{W}^2 M_Z^2}(|N_{31}|^2 - N_{21}|^2),
\end{equation}
where $g_\mathrm{A}^u=1/2$, $g_\mathrm{A}^d=g_\mathrm{A}^s=-1/2$, and $\Delta_q^N$ are the corresponding form factors for the nucleon $N=p,n$.

\subsection{Results}
\label{sec:SDFDM_model_results}

We perform a parameter scan in the 4-dimensional parameter space of the SDFDM model within the following ranges:
\begin{equation}
10~\si{GeV}< m_S, m_D < 1~\si{TeV},\quad
-2 < y_1, y_2 < 2.
\end{equation}
The experimental requirements are the same as those in Sec.~\ref{sec:STSDM_model_results}, except that the constraints from direct detection experiments on the SD scattering are also considered.
Here we adopt the the 90\% C.L. upper limit on $\sigma_p^\mathrm{SD}$ from PICO-60~\cite{PICO:2019vsc} and that on $\sigma_n^\mathrm{SI}$ from PandaX-4T~\cite{PandaX-4T:2021bab}.
However, we do not find any viable parameter point that can simultaneously explain the CDF $M_W$ anomaly and the correct DM relic density and avoid the direct detection constraints. This result can be understood as follows.  

Let us consider some typical parameter regions, where the stringent constraints from direct detection experiments are avoided. In order to escape from the SI constraint, some specific conditions leading to $g_{h\chi_1^0\chi_1^0} = 0$ have been discussed in Ref.~\cite{Xiang:2017yfs}. According to the low energy Higgs theorem~\cite{Ellis:1975ap}, we can read $g_{h\chi_1^0\chi_1}= \partial m_{\chi_1^0}(v)/\partial v$ from the Lagrangian
\begin{equation}
\mathcal{L}_{h\chi_1^0\chi_1^0}= \frac{1}{2} m_{\chi_1^0}(v) \chi_1^0\chi_1^0+ \frac{1}{2} \frac{\partial m_{\chi_1^0}(v)}{\partial v} h\chi_1^0\chi_1^0 +\mathcal{O}(h^2).
\end{equation}
$m_{\chi_1^0}(v)$ is determined by the characteristic equation $\mathrm{det}(M_\mathrm{N}-m_{\chi_1^0} \mathrm{I})=0$, which is equivalent to
\begin{equation}
m_{\chi_1^0}^3-m_Sm_{\chi_1^0}^2-\frac{1}{2}(2m_D^2+y_1^2v^2+y_2^2v^2)m_{\chi_1^0}+m_D(m_Dm_S+y_1y_2v^2)=0.
\label{masseuqation}
\end{equation}
Differentiating this equation and requiring $g_{h\chi_1^0\chi_1^0} = \partial m_{\chi_1^0}(v)/\partial v=0$, we get
\begin{equation}
m_{\chi_1^0}=\frac{2y_1y_2m_D}{y_1^2+y_2^2}.
\end{equation}
Applying this result in Eq.~\eqref{masseuqation}, we obtain a condition $y_1= y_2$ for $m_S>m_D>0$, and another condition
\begin{equation}
y_1=\frac{y_2}{m_S} \left(m_D\pm \sqrt{m_D^2-m_S^2}\right)
\end{equation}
for $m_D>m_S>0$.

For $m_S>m_D>0$, the condition $y_1=y_2$ leads to $m_{\chi_1^0}=m_D$ and $g_{h\chi_1^0\chi_1^0}=0$. This is the case in the custodial symmetry limit as discussed in Ref.~\cite{Cai:2016sjz}. We can define two $\mathrm{SU}(2)_\mathrm{R}$ doublets,
\begin{equation}
(\mathcal{D}^A)^i=
\begin{pmatrix}
D_1^i\\
D_2^i
\end{pmatrix},\quad
(\mathcal{H}^A)_i=
\begin{pmatrix}
H_i^\dagger\\
H_i
\end{pmatrix},
\end{equation}
and rewrite the Lagrangian of the NP sector with $y_1=y_2=y$ as
\begin{equation}
\mathcal{L}_\mathrm{NP} = i \mathcal{D}_A^\dagger\bar{\sigma}^\mu D_\mu \mathcal{D}^A-\frac{1}{2}m_D\epsilon_{AB}\epsilon_{ij}(\mathcal{D}^A)^i(\mathcal{D}^B)^j+y \epsilon_{AB} (\mathcal{H}^A)_i S(\mathcal{D}^B)^i+ \mathrm{H.c.}\,,
\end{equation}
which is invariant under a global $\mathrm{SU}(2)_\mathrm{L}\times \mathrm{SU}(2)_\mathrm{R}$ symmetry.
In this custodial symmetry limit, $\chi_1^0$ have equal components from the two doublets with opposite hypercharges, resulting in a vanishing $Z\chi_1^0\chi_1^0$ coupling and hence vanishing SD scattering cross sections. Therefore, all the constraints from direct detection are evaded.
However, when $y_1=y_2$, the $T$ and $U$ parameters exactly vanish due to the custodial symmetry~\cite{Peskin:1991sw}. As a result, the parameter points with $y_1\sim y_2$ cannot induce a large $T$ or $U$ parameter, which is required for explaining the CDF $M_W$ anomaly.

\begin{figure}
\centering
\includegraphics[width=0.48\textwidth]{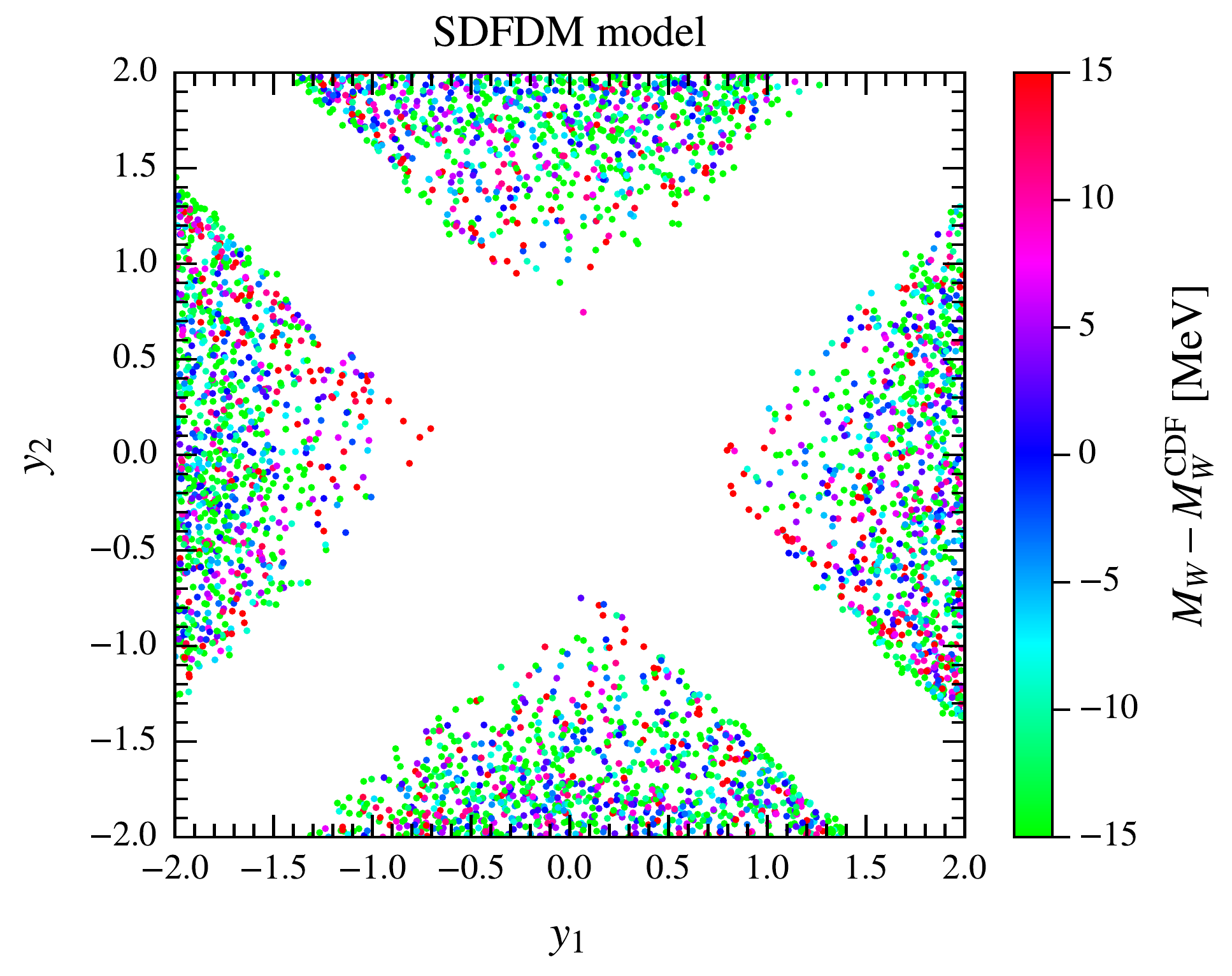}
\caption{Parameter points accounting for the CDF $M_W$ anomaly projected in the $y_1$-$y_2$ plane for the SDFDM model. The colors indicate the difference between the predicted $M_W$ and the central value of the CDF measurement $M_W^{\mathrm{CDF}}$. Note that these parameter points are not required to satisfy the DM relic density and direct detection constraints.
}\label{fig:SDFDM1}
\end{figure}

In Fig.~\ref{fig:SDFDM1}, we show the parameter points in the $y_1$-$y_2$ plane accounting for the $M_W$ anomaly, and do not find points in the region with $y_1 \sim y_2$. There is also no point with $y_1\sim -y_2$, due to another custodial symmetry limit $y_1 = - y_2$, which is corresponding to the $\mathrm{SU}(2)_\mathrm{R}$ doublet defined by $(\mathcal{H}^A)_i=(-H_i^\dagger, H_i)$ instead.
In this limit, we also have $T = U = 0$ and $g_{Z\chi_1^0\chi_1^0}=0$.

For $m_D>m_S>0$, the condition $y_1/y_2=(m_D\pm \sqrt{m_D^2-m_S^2})/m_S$ leads to $m_{\chi_1^0}=m_S$ and $g_{h\chi_1^0\chi_1^0}=0$.
For small $y_1$ and $y_2$, $\chi_1^0$ is dominated by the singlet component. The correct DM relic density could be derived for relatively large $y_1$ and $y_2$, which lead to moderate doublet components in $\chi_1^0$. However, increasing the doublet components would lift up the $Z\chi_1^0\chi_1^0$ coupling, more easily to be excluded by the SD direct detection constraints. We fix $y_1=y_2 (m_D\pm \sqrt{m_D^2-m_S^2})/m_S$ and perform a scan with three free parameters, $m_S$, $m_D$, and $y_2$. No point satisfying all the experimental requirements is found in this scan.

\begin{figure}
\centering
\subfigure[~$m_{\chi_1^0}-\sigma^\mathrm{SI}_N$ plane\label{fig:SDFDM2:SI}]{\includegraphics[width=0.48\textwidth]{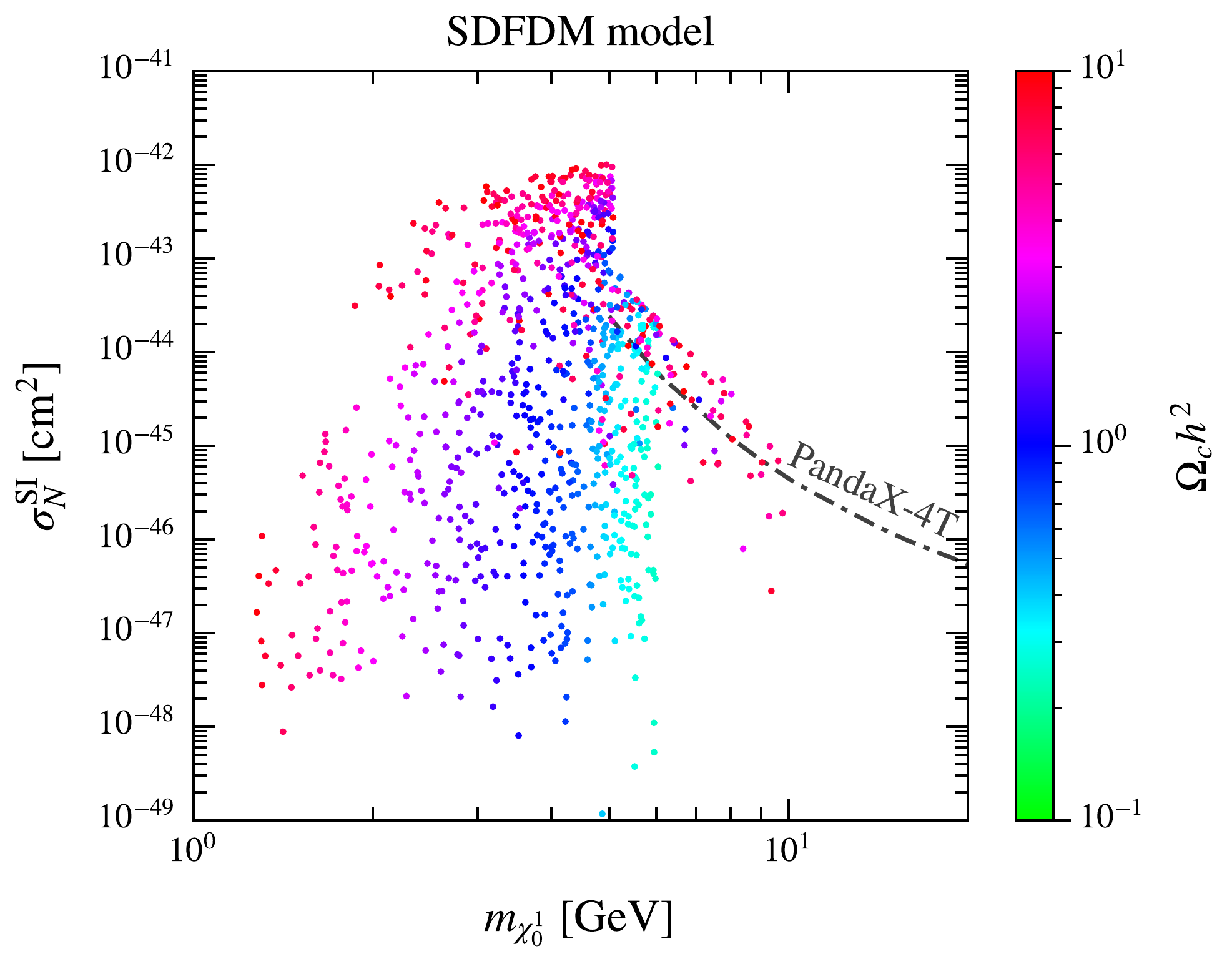}}
\subfigure[~$m_{\chi_1^0}-\sigma^\mathrm{SD}_p$ plane\label{fig:SDFDM2:SD}]{\includegraphics[width=0.48\textwidth]{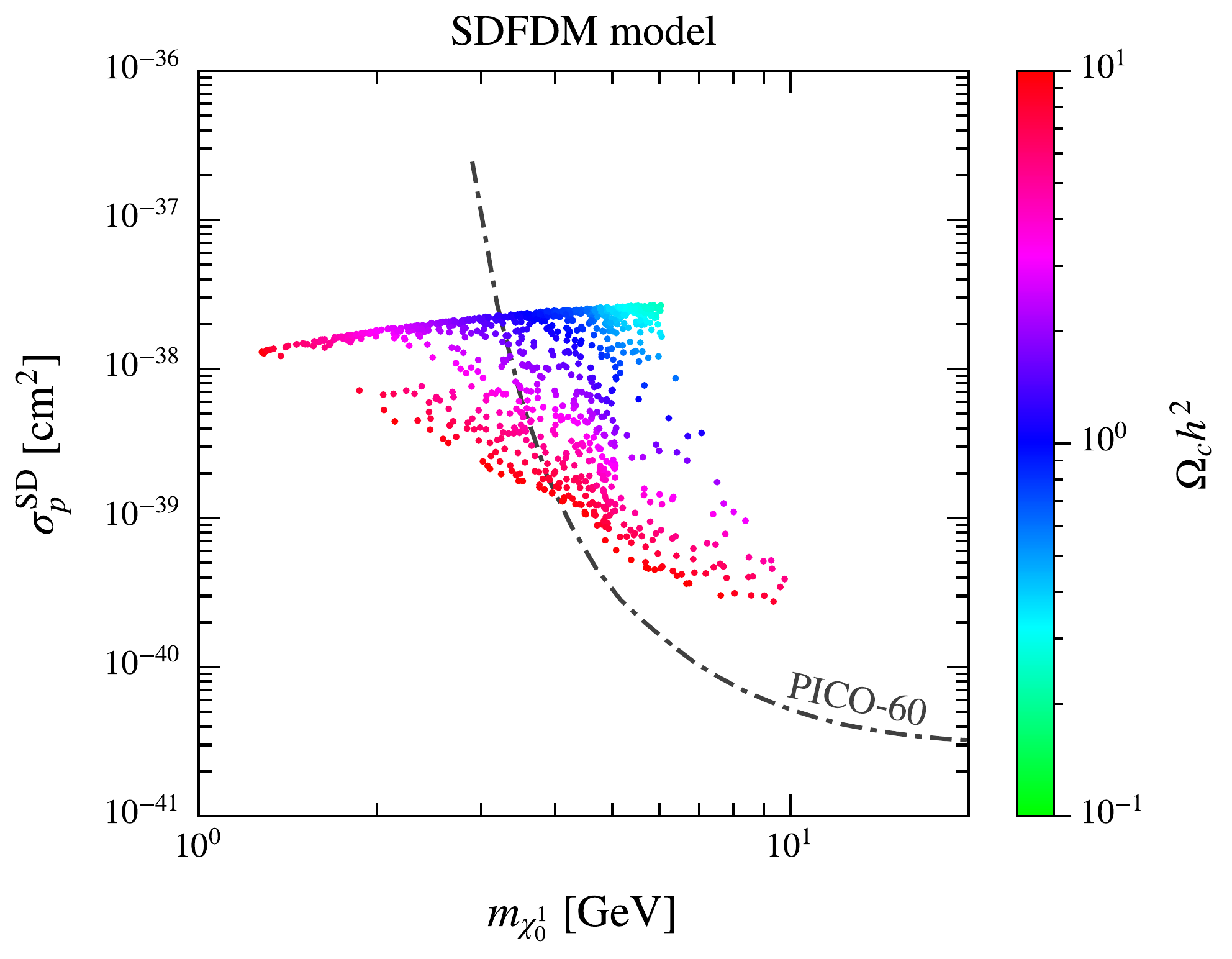}}
\caption{Parameter points accounting for the CDF $M_W$ anomaly projected in the $m_{\chi_1^0}$-$\sigma^\mathrm{SI}_N$ (a) and $m_{\chi_1^0}$-$\sigma^\mathrm{SD}_p$ (b) planes. The colors indicate the DM thermal relic density. The 90\% C.L. upper limit on $\sigma^\mathrm{SI}_N$ from the PandaX-4T experiment~\cite{PandaX-4T:2021bab} is indicated in the left panel, while the 90\% C.L. upper limit on $\sigma^\mathrm{SD}_p$ from the PICO-60 experiment~\cite{PICO:2019vsc} is demonstrated in the right panel. Note that the parameter points leading to the DM-nucleon scattering cross sections much larger than the limits from the direct detection experiments are not shown here.}
\label{fig:SDFDM2}
\end{figure}

The direct detection constraints become rather weak or even vanish when the DM particle mass is below a few GeV, due to the kinematic thresholds in the experiments. It can be observed that Eq.~\eqref{masseuqation} has a solution of $m_{\chi_1^0}=0$ for $y_1=-m_D m_S/(y_2 v^2)$. Therefore, for $y_1\sim -m_D m_S/(y_2 v^2)$, very light $\chi_1^0$ with $m_{\chi_1^0}\sim \mathcal{O}(1)~\si{GeV}$ could evade the constraints from direct detection experiments. There could be a large splitting in the mass spectrum of the NP fermions, which is useful for generating large oblique parameters. If we only impose the requirements of the $M_W$ anomaly and the direct detection constraints, all the parameter points allowed in the scan correspond to small $m_{\chi_1^0}$. However, since $m_{\chi_1^0}$ is too small, the $\chi_1^0 \chi_1^0$ annihilation channels are rather limited, leading to highly suppressed annihilation cross sections and overproduction of dark matter at the freeze-out epoch. This can be seen in Fig.~\ref{fig:SDFDM2}, where the parameter points accounting for the $M_W$ anomaly are shown in the $m_{\chi_1^0}$-$\sigma^\mathrm{SI}_N$ and $m_{\chi_1^0}$-$\sigma^\mathrm{SD}_p$ planes, compared with the direct detection upper limits. The parameter points with very small $m_{\chi_1^0}$ allowed by the direct detection constraints generally predict a large relic density, and no parameter point satisfy the relic density observation.

\section{Conclusions}
\label{sec:conclusions}

In this paper, we study the interpretation of the recent CDF $M_W$ anomaly in the DM models involving extra EW multiplets. We consider both the STSDM and SDFDM models, where new scalar and fermionic multiplets are introduced, respectively. It is obvious that these extra EW multiplets would affect the $W$ boson mass and other EW precision observables through loop corrections. We perform numerical scans in the parameter space of the two models, and attempt to find the parameter points that can simultaneously explain the CDF $M_W$ anomaly and satisfy other EW precision tests and the DM constraints. 

In order to generate the measured $M_W$ deviation from the SM prediction, there should be some moderate EW oblique parameters contributed by the NP particles. If such contributions are loop-induced, the corresponding NP particles are required to have masses below a few hundred GeV, and the mass splitting in the NP mass spectrum should be relatively large. 

In the STSDM model containing an inert real scalar singlet and an inert complex scalar triplet, we find that many parameter points satisfying all the DM constraints can interpret the CDF $M_W$ anomaly.
%Note that the correct DM relic density sets stringent constraint on the selected parameter points accounting for the $M_W$ anomaly. This is because that the light DM particles participating the weak interactions would have a relative large annihilation cross section, which leads to a too small thermal relic density.
We observe that the mass range of the DM candidate required by the $M_W$ anomaly is strongly constrained by direct detection experiments, generally leading to small scalar couplings and overproduction of dark matter.
In order to avoid this problem, exceptional parameter regions with significant effects of coannihilation, resonantly annihilation, and annihilation to the forbidden channels are typically required to give the correct relic density.
Almost all viable parameter points can be classified in these exceptional cases. 

In the SDFDM model containing a Weyl singlet and two Weyl doublets, we do not find any parameter point simultaneously satisfying the requirements of the $M_W$ anomaly and DM phenomenology. As a check, we focus on some parameter regions with a vanishing coupling $g_{h\chi_1^0\chi_1}$ or very small $m_{\chi_1^0} \sim \mathcal{O}(1)~\si{GeV}$, in order to avoid the stringent constraints from the SI direct detection. We find that the parameter regions with $g_{h\chi_1^0\chi_1}\sim 0$ cannot explain the $M_W$ anomaly or are excluded by the constraints from the SD direct detection, while the parameter regions with small $m_{\chi_1^0}$ cannot give the correct DM relic density. Compared with scalar DM models, fermionic DM models generally have less couplings and less free parameters. Moreover, fermionic DM could also have SD scattering off nucleons in direct detection experiments.
These differences make it more difficult for fermionic DM models to simultaneously satisfy all the experimental requirements.

The scenario accounting for the $M_W$ anomaly discussed in this paper can be further tested by future experiments. Since the NP particles interacting with the EW gauge bosons and Higgs bosons are quite light in this scenario, it is expected that they would have significant production rates at high energy colliders. Furthermore, the NP particles with EW interactions would also modify
the production and decay rates of the SM particles via loop corrections. These signatures and effects could be well explored in future collider experiments.

\begin{acknowledgments}
The works of XJB and ZHY are supported by the National Natural Science Foundation of China under Grant Nos.~12175248 and 11805288, respectively.
The work of JWW is supported by the research grant ``the Dark Universe: A Synergic Multi-messenger Approach'' number 2017X7X85K under the program PRIN 2017 funded by the Ministero dell'Istruzione, Universit\`{a} e della Ricerca (MIUR).
\end{acknowledgments}

\bibliographystyle{utphys}
\bibliography{ref}

\end{document}